\documentstyle[12pt,aaspp4]{article}
\begin{document}

\title {Unusual Burst Emission from the New Soft Gamma Repeater SGR1627-41}
\author{
E.P.Mazets\altaffilmark{1,2},
R.L.Aptekar\altaffilmark{1},
P.S.Butterworth\altaffilmark{3,4},
T.L.Cline\altaffilmark{3},
D.D.Frederiks\altaffilmark{1},
S.V.Golenetskii\altaffilmark{1},
K.Hurley\altaffilmark{5,6},
V.N.Il'inskii\altaffilmark{1}
}

\altaffiltext{1}{Ioffe Physico-Technical Institute, St.Petersburg, 194021,
Russia.}
\altaffiltext{2}{Mazets@pop.ioffe.rssi.ru}
\altaffiltext{3}{Goddard Space Flight Center, Greenbelt, MD 20771, USA.}
\altaffiltext{4}{Butterworth@lheavx.gsfc.nasa.gov}
\altaffiltext{5}{Space Science Laboratory, University of California,
Berkeley, CA 94720-7450, USA.}
\altaffiltext{6}{KHurley@sunspot.ssl.berkeley.edu}

\begin{abstract}
In June-July~1998 the Konus-Wind burst spectrometer observed a series of
bursts
from the new soft gamma repeater SGR~1627-41. Time histories and energy
spectra
of the bursts have been studied, revealing fluences and peak fluxes in
the ranges of $3\cdot 10^{-7}$ - $7.5\cdot 10^{-6}~erg~cm^{-2}$ and
$10^{-5}$ - $10^{-4}~erg~cm^{-2}~s^{-1}$ respectively.
One event, 18~June 6153.5s~UT stands out dramatically from this series.
Its fluence is $\sim 7\cdot 10^{-4}~erg~cm^{-2}$ and peak flux
$\sim 2\cdot 10^{-2}~erg~cm^{-2}~s^{-1}$. These values from a
source at a distance of 5.8~kpc yield an energy output of
$\sim 3\cdot 10^{42}~erg$ and maximum luminosity of $\sim 8\cdot
10^{43}~erg~s^{-1}$,
similar to the values for the famous March~5, 1979 and August~27, 1998
events. In terms of energy, this event is another giant outburst
seen in a third SGR! However, this very energetic burst differs
significantly
from the other giant outbursts. It exhibits no separate initial pulse with a
fast
rise time, no extended tail, and no pulsations. It is rather similar to
ordinary
repeated bursts but is a few hundred times stronger in intensity. According
to the
magnetar model by Thompson and Duncan (1995) such a burst may be initiated by a
strong
starquake when a crust fracture propagates over the whole surface of a
neutron
star.
\end{abstract}


\section{Introduction}

Repeating soft gamma-ray bursts were discovered 20~years ago (Golenetskii,
Ilyinskii, \& Mazets, 1984; Mazets, Golenetskii, \& Guryan, 1979;
Atteia et al. 1987). For a long time only three soft gamma repeaters
were known (Norris, Hertz, \& Wood 1991) suggesting a rarity
of this class of astrophysical objects (Kouveliotou et al. 1998). Two of
them,
SGR~1806-20 and SGR~1900+14, have exhibited reactivation phases after
periods
of long silence. Precise localizations and a search for optical counterparts
revealed that all three SGRs are associated with rather young supernova
remnants (Cline et al. 1982; Kulkarni et al. 1994; Hurley et al. 1999a)
favoring a suggestion that the SGR's are neutron stars. Quiescent soft X-ray
sources and emission periodicity were discovered associated with SGR~1806-20
(Murakami et al. 1994), SGR1900+14 (Hurley et al. 1999b) and SGR~0526-66
(Rothschild, Kulkarni \& Lingenfelter 1994). A spectacular huge periodic
flare
on August~27,1998 which came from SGR~1900+14 turned out to be  strikingly
similar to the famous event on March~5,1979. It demonstrated that such
giant outbursts are an intrinsic but less common characteristic of SGRs
(Cline, Mazets \& Golenetskii 1998). Observational data accumulated so far
on SGRs have found their most complete explanation in the magnetar model by
Thompson and Duncan (Thomson \& Duncan 1995) which proposes that SGRs are
young slowly
rotating neutron stars with superstrong magnetic fields of $\sim 10^{15}~G$.

In June 1998, BATSE announced an observation of a fourth soft gamma repeater
SGR~1627-41 (Kouveliotou et al. 1998 ) confirmed by Ulysses (Hurley, et al.
1998a), Beppo Sax (Feroci, et al. 1998), RXTE (Smith \& Levine 1998), and
Konus-Wind (Hurley et al. 1998b) data. This SGR was precisely localized by
IPN/Beppo Sax (Hurley et al. 1999d; Woods et al. 1999). Its position
coincides
with the SNR~G337.0-0.1. Some evidence was obtained for a possible
periodicity
of 6.7~s (Dieters et al. 1998).

In this letter we report temporal and spectral properties of SGR~1627-41
as well as an unusual behavior of this source recorded in Konus-Wind
observations.

\section{Observations}

The cosmic gamma-ray burst spectrometer Konus aboard the GGS~WIND spacecraft
observed 34 bursts from the new SGR in the period June~17-July~12,1998.
Some events were too weak to trigger the instrument. They are recorded in
a background mode with a time resolution of 2.94~s which is too coarse to
study
processes a small fraction of a second in duration. The bursts from the new
source were strongly bunched in time. In two days, June,17 and 18, 29 bursts
were observed. This high rate created an additional problem. The information
on a burst which triggered the instrument is read by the S/C TM-system over
a period of
$\sim$ one hour. If another burst occurs in this time interval, it can't
trigger
the instrument and is recorded only among housekeeping data with 3.86~s
resolution.
As a result, we obtained triggered records of only 13 bursts, which
nevertheless give a good idea
of the temporal and spectral behaviour of this SGR. Regrettably, as we can
see
from
the untriggered data, several very interesting events were not caught with
high time resolution.

In Figure~1 we show some examples of time histories of bursts recorded with
a time resolution of 2~ms at photon energy $E_{\gamma} > 15$~keV.
One can see that temporal structures are rather complex with rise
and fall times of only a few milliseconds. Energy spectra, as expected,
are soft with an exponential cutoff, $kT\sim 25~keV$. Spectral variability
is apparent in the course of many bursts. For example, time histories
G1 and G2 of the burst on June~18, 16229.0s~UT recorded in the low energy
window G1(15 -- 55~keV) and the middle one G2(55 -- 250~keV) as well as
the hardness ratio G2/G1 are shown in Figure~2. Detailed energy spectra
for this event, accumulated after triggered time $T_o$ in subsequent 64~ms
long time intervals, are presented in Figure~3. These photon spectra can be
fitted by the expression $dN/dE \sim E^{-0.5}~e^{-E/kT}$.
A weak tail of the bursts exhibits a fading power law spectrum with an index
of $-2.8$. Data obtained for 13 events are sampled in Table~1 which contains
values of kT, fluences, and peak fluxes; as well as the energy output and
maximum
luminosity of the source assuming a 5.8~kpc distance (Case \& Bhattacharya
1998).

One event from this Table, GRB~980618a 6153.5s~UT, stands out dramatically
in intensity compared to the other bursts. The intensity of this event
is so high  that count rates in the low and middle energy windows
approached saturation level. High count rates appear in the high energy
window
G3(250 -- 1000~keV) due to the pile up of light pulses in the scintillator
and
photomultiplier. The instrument response to very high fluxes of various
incident photon spectra were studied thoroughly in a laboratory simulation
using a spare unit (Mazets, et al. 1999). This enables a reliable
deconvolution of the Konus-Wind input fluxes. The time history of the giant
outburst, after correction for dead time and pile up effects,
is shown in Figure~4. Possible
errors in the peak flux region can be represented by a scale factor of 0.5
-- 1.5.

Figure~5 presents energy loss spectra accumulated during four adjacent
64~ms long intervals and one of 256~ms. The very hard spectrum in the
64-128~ms
interval is partially the result of pile up of light flashes in the
NaI crystal. From laboratory testing data however, it is evident that in
this case the actual incident photon spectrum is also much harder than the
spectrum
for the first interval $T-T_o$~=~0-64~ms. Its kT can be as high as
100-150~keV
whereas unaffected spectra A, D, E correspond to kT: 50, 50, and 35~keV
respectively.

\section{Discussion and conclusion}

The fluence and peak flux of this event exceed by several hundred times the
values for other bursts of the series.  At an assumed
distance to the source of 5.8~kpc, they indicate an energy output of
$\sim 3\cdot 10^{42}~erg$
and a peak luminosity of $\sim 8\cdot 10^{43}~erg~s^{-1}$. These quantities
approach the huge energy releases and maximum luminosities for the giant
outbursts on
March~5,1979 and August~27,1998 (Hurley, et al. 1999c). From this point
of view, the burst on June~18,1998 6153.5~s~UT is also a giant outburst seen
in a third SGR! But this burst differs strikingly from them.
It exhibits no fast rise time, no long tail, and no pulsations. It is
similar to ordinary repeating bursts from SGR~1627-41, but much
stronger.

It is now widely accepted that an SGR's normal activity is a result
of starquakes on neutron stars leading to a liberation of a large amount
of the energy contained in superstrong magnetic fields of $\sim 10^{15}~G$
(Thompson \& Duncan 1995). These authors have proposed that giant outbursts
like the March~5 event are created by large scale reconnection instabilities
of a huge stellar magnetic field. The more frequent weaker repeating bursts
are
generated during fractures of the neutron star crust. It was suggested that
the energy of starquake-related bursts (i.e. their observed fluences)
may depend on the size/length
of fractures (Duncan, 1998; Golitsin, 1998). In such behavior an upper limit
for the burst energy output will be determined by the maximum length of
fractures
crossing over the whole surface of the neutron star. It seems that the burst
on
June~18,1998 may be the first observed example of an event close to such
an upper limit.

It is expected that rise times of repeated bursts must relate to their
fluence S as $\tau \sim S^{1/3}$ (Golitsin, 1998). For weak bursts of
(3-6) $10^{-7}~erg~cm^{-2}$ rise times are $\sim 10~ms$. For the giant
event with $S\sim 2\cdot 10^{-4} erg cm^{-2}$ the rise time is
$\sim 10^{2}$ ms, in accord with the prediction.

There is some evidence that such energetic bursts may be quite common.
Among bursts detected with Konus-Wind in untriggered background mode there
are at least two events, June~18~14360~s~UT and June~18~14661~s~UT, with
very high fluences. If their durations were short compared to the time
resolution of 3.86~s, then their intensities corrected for
dead time could be comparable with the energy fluence
of the strong June~18 burst under discussion.

These results provide an important confirmation of the SGR
starquake model predictions.

\acknowledgments
This work on the Russian side was supported by RSA contract.

\newpage
\begin{deluxetable}{lcccccc}
\tablecaption{Main characteristics of observed bursts}
\tablehead{
\colhead{GRB}& \colhead{$T_o(UT)$}& \colhead{kT}& \colhead{S}& \colhead{$F_{max}$}& \colhead{Q}& \colhead{L}\\
\colhead{}& \colhead{$s$}& \colhead{$keV$}& \colhead{$erg~cm^{-2}$}& \colhead{$erg~cm^{-2}~s^{-1}$}& \colhead{$erg$}& \colhead{$erg~s^{-1}$}
}    

\startdata
980617a&   71912.0&   	22&   	$3.3\cdot 10^{-6}$&	$3.7\cdot 10^{-5}$&   $1.3\cdot 10^{40}$&   $1.5\cdot 10^{41}$\\
980617b&   75880.0&     14&     $1.2\cdot 10^{-6}$&    $3.3\cdot 10^{-5}$&   $4.8\cdot 10^{39}$&   $1.3\cdot 10^{41
}$\\
980617c&   82449.0&     18&     $7.0\cdot 10^{-7}$&    $1.0\cdot 10^{-5}$&   $2.8\cdot 10^{39}$&   $4.0\cdot 10^{40}$\\
980618a&    6153.5&    $\sim$150&     $7.0\cdot 10^{-4}$&    $2.0\cdot 10^{-2}$&   $3.0\cdot 10^{42}$&   $8.0\cdot 10^{43}$\\  
980618b&   12316.5&     16&     $6.5\cdot 10^{-7}$&    $9.0\cdot 10^{-6}$&   $2.6\cdot 10^{39}$&   $3.6\cdot 10^{40}$\\
980618c&   16229.0&     23&     $7.5\cdot 10^{-6}$&    $1.0\cdot 10^{-4}$&   $3.0\cdot 10^{40}$&   $4.0\cdot 10^{41}$\\
980618d&   22593.1&     20&     $3.0\cdot 10^{-7}$&    $1.0\cdot 10^{-5}$&   $1.2\cdot 10^{39}$&   $4.0\cdot 10^{40}$\\
980618e&   59884.7&     18&     $5.5\cdot 10^{-7}$&    $7.0\cdot 10^{-6}$&   $2.2\cdot 10^{39}$&   $2.8\cdot 10^{40}$\\
980622a&   48596.3&     24&     $4.5\cdot 10^{-6}$&    $5.0\cdot 10^{-5}$&   $1.8\cdot 10^{40}$&   $2.0\cdot 10^{41}$\\
980622b&   68199.1&     17&     $6.0\cdot 10^{-7}$&    $3.0\cdot 10^{-5}$&   $2.4\cdot 10^{39}$&   $1.2\cdot 10^{41}$\\
980625&    39379.6&     24&     $4.3\cdot 10^{-6}$&    $5.0\cdot 10^{-5}$&   $1.7\cdot 10^{40}$&   $2.0\cdot 10^{41}$\\
980629&    26706.8&     17&     $4.7\cdot 10^{-7}$&    $1.2\cdot 10^{-5}$&   $1.9\cdot 10^{39}$&   $4.8\cdot 10^{40}$\\
980712&    78640.1&     16&     $5.4\cdot 10^{-7}$&    $2.0\cdot 10^{-5}$&   $2.2\cdot 10^{39}$&   $8.0\cdot 10^{40}$\\
\enddata
\end{deluxetable}

\clearpage

\figcaption{Time histories of 5 bursts from SGR~1627-41.
Time resolution 2~ms, photon energy $E_{\gamma}>15$~keV. Complex temporal
structures are evident.}

\figcaption{The burst on June~18, 16229s UT. Time histories taken
in two energy windows G1(15-55 keV) and G2(55-250 keV). A fast strong
spectral variability manifests itself in a hardness ratio G1/G2 profile. 
In time intervals denoted A, B, C, D, E energy spectra were accumulated.}

\figcaption{The burst on June~18, 16229s~UT. Subsequent energy spectra
A, B, C, D, E are shifted for clarity by factors of 1; 1/5; 1/50;
1/250; 1/250 respectively. The spectra A-D show a variation of parameter
kT: $22.0\pm 0.7$; $18.0\pm 0.7$; $28\pm 1.5$; $23\pm 1$~keV. 
The last spectrum is fitted by a power law with an index 
$\alpha = - 2.8\pm 0.3$.}

\figcaption{The time history of the giant burst on June~18, 6153s~UT
corrected for dead time. Photon energy $E_{\gamma}>15$~keV. The rise time
is about 100~ms.}

\figcaption{Energy loss spectra measured in adjacent time intervals
A, B, C, D, E and shifted for clarity by factors of 1/5; 10; 1; 1/25;1/125
respectively. Spectra B, C are strongly affected by pile up effects, 
especially spectrum B near the peak of the burst.}

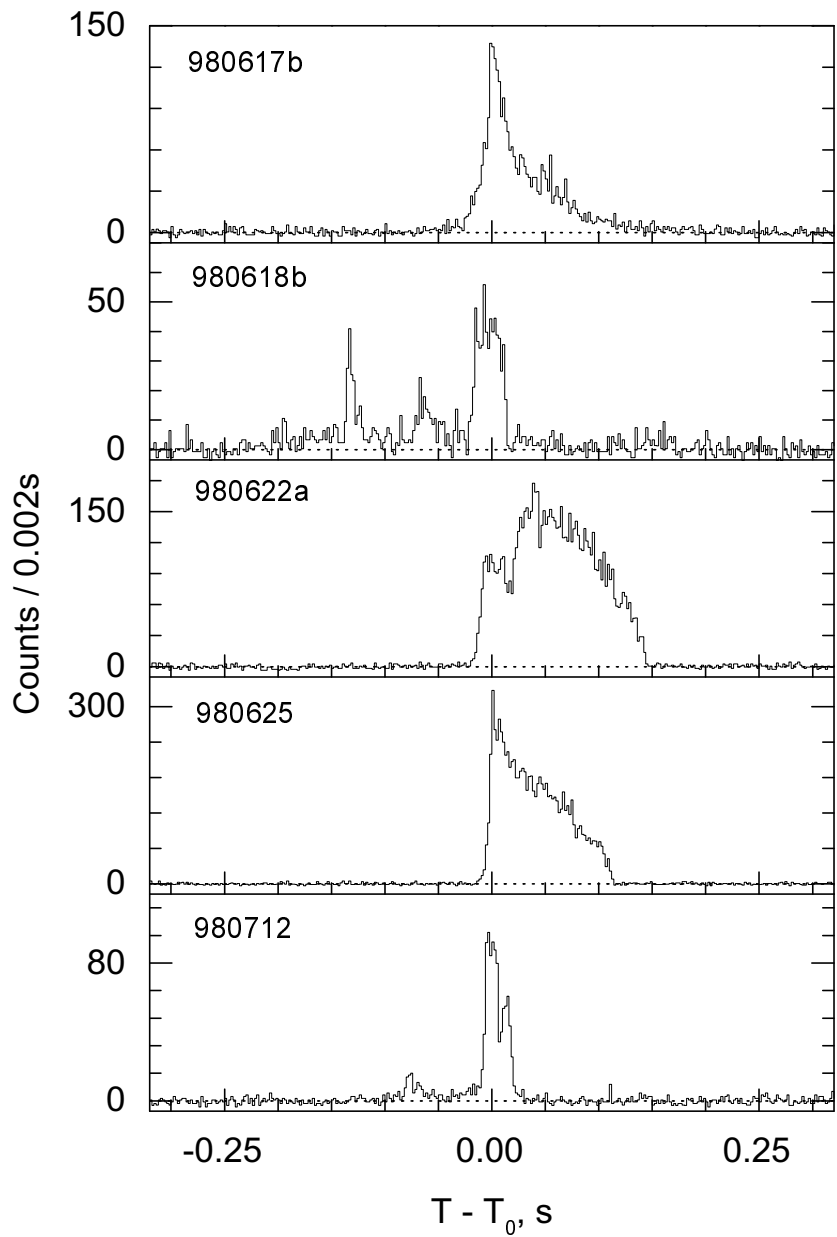
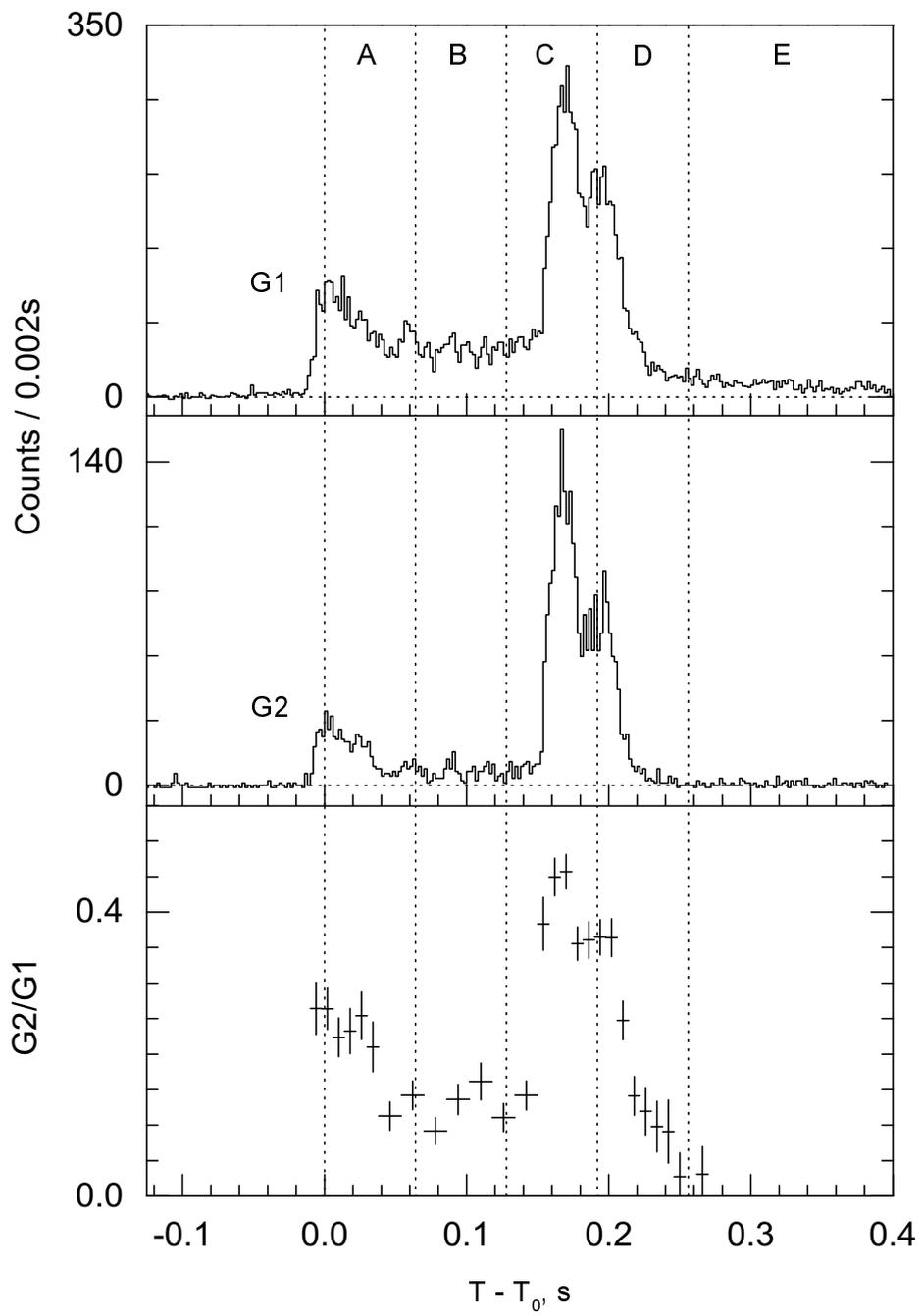
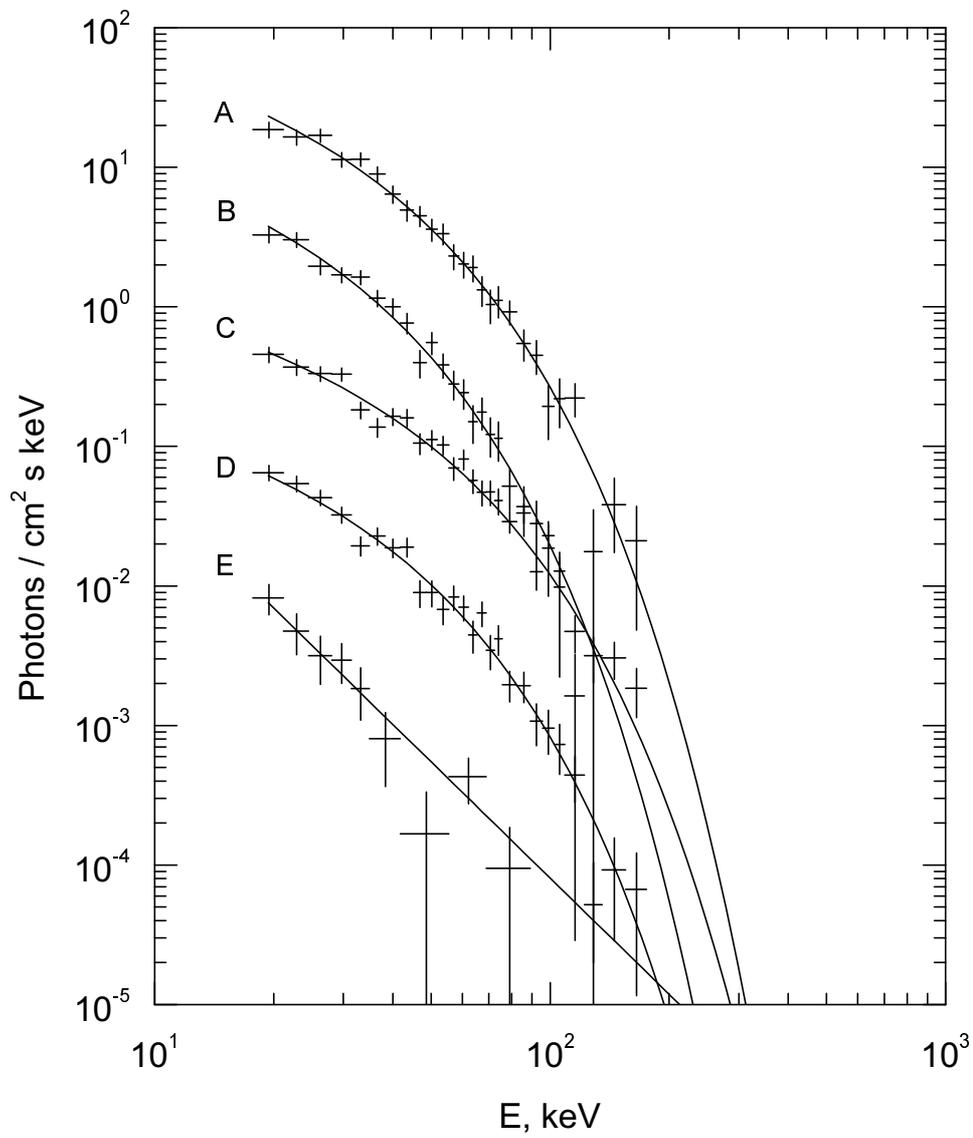
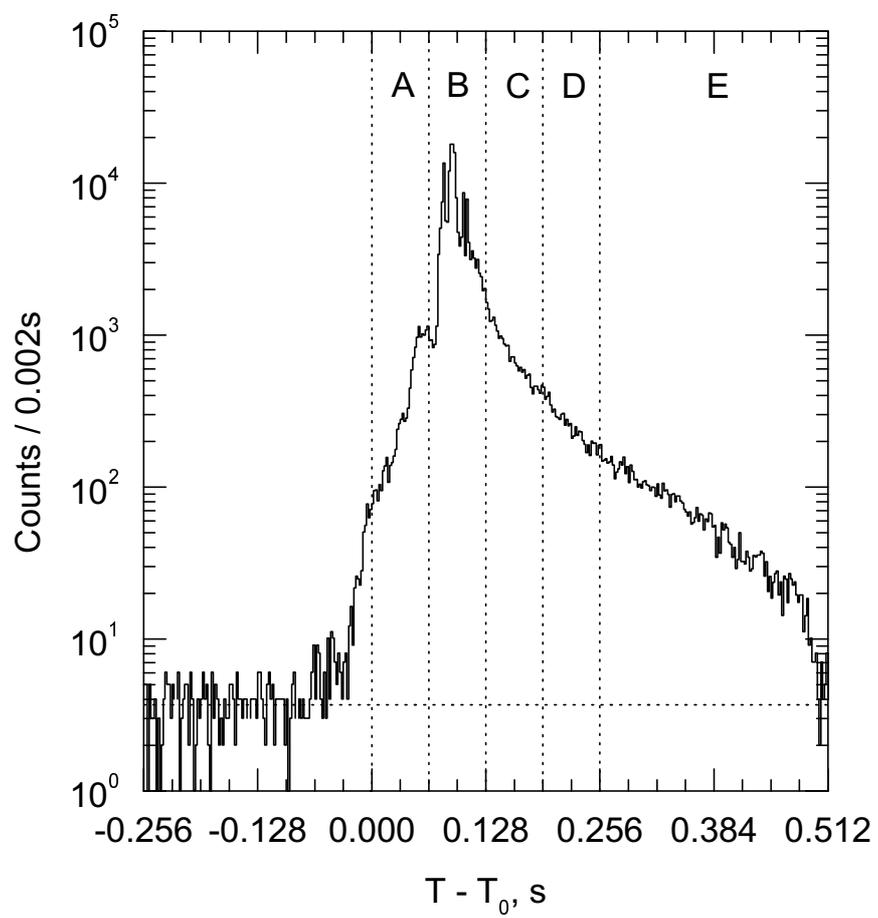
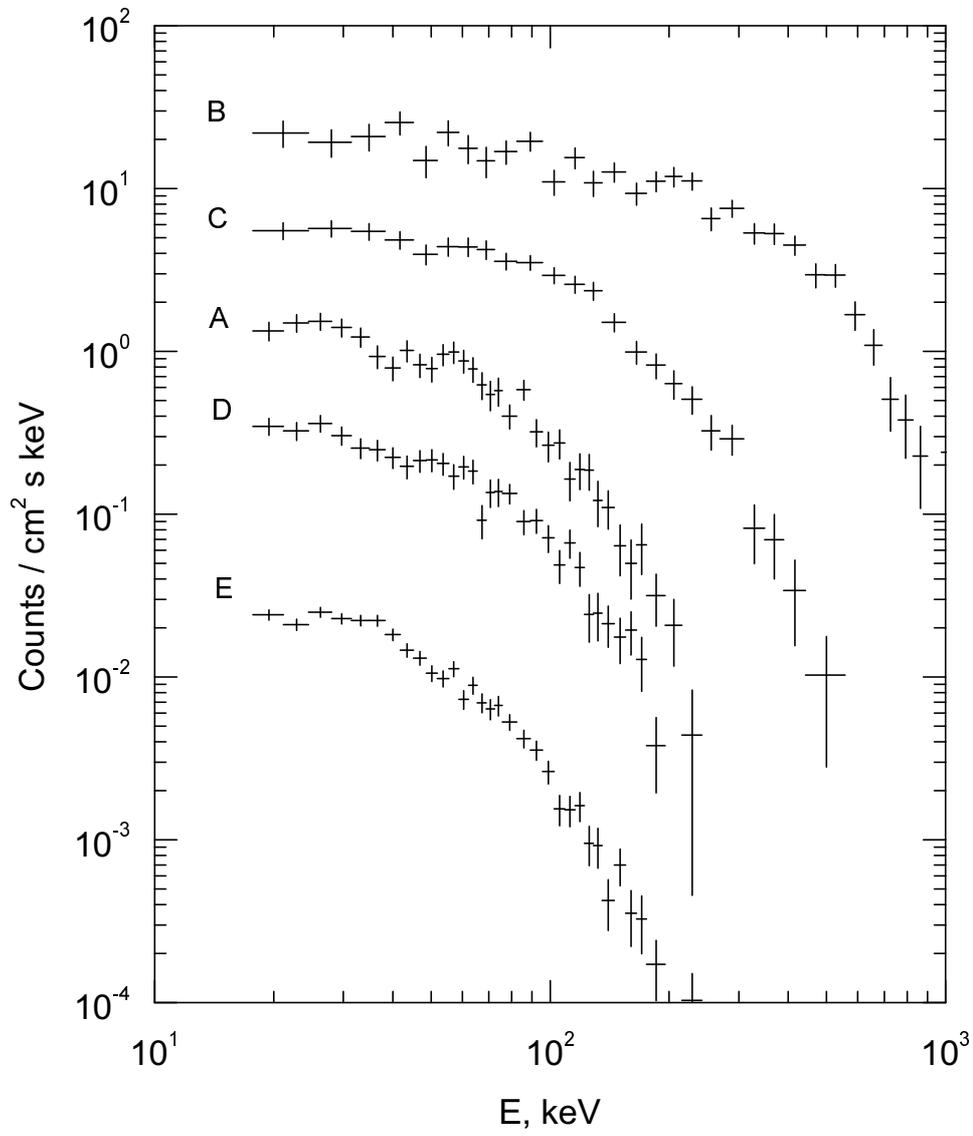
\end{document}